\documentclass{amsart}
\usepackage{amssymb, natbib}

\RequirePackage{ifthen, url}

\newtheorem{theorem}{Theorem}
\newtheorem{lemma}{Lemma}
\newtheorem{Claim}{Claim}
\newtheorem{proposition}{Proposition}
\newtheorem{corollary}{Corollary}

\theoremstyle{definition}

\newtheorem{example}{Example}

\theoremstyle{remark}

\newtheorem{conjecture}{Conjecture}

\newcommand{\rslt}[1]{\ifthenelse{\ref{lem:#1} > 0}{lemma \ref{lem:#1}}%
{%
\ifthenelse{\ref{prp:#1} > 0}{proposition \ref{prp:#1}}%
{%
\ifthenelse{\ref{thm:#1} > 0}{theorem \ref{thm:#1}}%
{%
\ifthenelse{\ref{cor:#1} > 0}{corollary \ref{cor:#1}}%
{%
\ifthenelse{\ref{clm:#1} > 0}{claim \ref{clm:#1}}%
{%
\ifthenelse{\ref{cnj:#1} > 0}{conjecture \ref{cnj:#1}}%
{%
\ifthenelse{\ref{xmpl:#1} > 0}{example \ref{xmpl:#1}}{\textbf{result
    ??} \typeout{Result #1 is not defined.}}%
}%
}%
}%
}%
}%
}%
}

\newcommand{\Rslt}[1]{\ifthenelse{\ref{lem:#1} > 0}{Lemma \ref{lem:#1}}%
{%
\ifthenelse{\ref{prp:#1} > 0}{Proposition \ref{prp:#1}}%
{%
\ifthenelse{\ref{thm:#1} > 0}{Theorem \ref{thm:#1}}%
{%
\ifthenelse{\ref{cor:#1} > 0}{Corollary \ref{cor:#1}}
{%
\ifthenelse{\ref{clm:#1} > 0}{Claim \ref{clm:#1}}%
{%
\ifthenelse{\ref{cnj:#1} > 0}{Conjecture \ref{cnj:#1}}%
{%
\ifthenelse{\ref{xmpl:#1} > 0}{Example \ref{xmpl:#1}}{\textbf{Result
    ??} \typeout{Result #1 is not defined.}}%
}%
}%
}%
}%
}%
}%
}

\newcommand{\eqnref}[1]{\ref{eqn:#1}}
\newcommand{\rsltref}[2]{\ref{#2:#1}} 
\newcommand{\secref}[1]{\ref{sec:#1}}

\newcommand{\lema}[2]{\begin{lemma} #2 \label{lem:#1} \end{lemma}}
\newcommand{\lem}[1]{{l}emma \ref{lem:#1}}

\newcommand{\claim}[2]{\begin{Claim} #2 \label{clm:#1} \end{Claim}}

\newcommand{\propo}[2]{\begin{proposition} #2 \label{prp:#1} \end{proposition}}
\newcommand{\prp}[1]{{p}roposition \ref{prp:#1}}

\newcommand{\exmpl}[2]{\begin{example} #2 \label{xmpl:#1} \end{example}}
\newcommand{\xmpl}[1]{{e}xample \ref{xmpl:#1}}

\newcommand{\display}[2]{\begin{equation} #2 \label{eqn:#1} \end{equation}}
\newcommand{\eqn}[1]{(\ref{eqn:#1})}

\newcommand{\Section}[2]{\section{#2\label{sec:#1}}}
\newcommand{\Subsection}[2]{\subsection{#2\label{sec:#1}}}
\renewcommand{\sec}[1]{{s}ection \ref{sec:#1}}

\newenvironment{enum} % Top-level enumeration environment
{\begin{list}{\makebox[\labelwidth][l]{(\arabic{enumi})}}{\usecounter{enumi}}
\setcounter{enumi}{\value{equation}}}
{\setcounter{equation}{\value{enumi}} \end{list}}

\newcommand{\meti}[2]{\item #2 \label{eqn:#1}} % \item with label

\newcommand{\abbrevEnvir}{
\expandafter\newcommand\expandafter{\csname bi\endcsname}{\begin{itemize}} 
\expandafter\newcommand\expandafter{\csname ei\endcsname}{\end{itemize}}
\expandafter\newcommand\expandafter{\csname be\endcsname}{\begin{enumerate}} 
\expandafter\newcommand\expandafter{\csname ee\endcsname}{\end{enumerate}}
\expandafter\newcommand\expandafter{\csname bc\endcsname}{\begin{center}} 
\expandafter\newcommand\expandafter{\csname ec\endcsname}{\end{center}}
}

\newcommand{\from}{\colon}

\renewcommand{\Re}{\mathbb{R}}

\author[F.~Borhani]{Fatemeh Borhani}
\email{fatemeborhani@gmail.com}
\author[E.~J.~Green]{Edward J.~Green}
\address{Department of Economics, Pennsylvania State University}
\email{eug2@psu.edu}
\title[Identifying cognitive bias]
{Identifying the occurrence or non occurrence\\ of cognitive bias in
  situations\\ resembling the \uppercase{M}onty \uppercase{H}all problem}
\date{2018.02.10}
\thanks{JEL Subject classes: D01, D03, D81}
\thanks{The authors thank Nageeb Ali for discussion and comments that
  have enhanced both the substance and exposition of the
  paper. Borhani's research was partly conducted as a visiting faculty
  member of the University of Pittsburgh.}

\newcommand{\mand}{\text{\ and\ }}

\abbrevEnvir

\newcommand{\hide}[1]{\relax}

\newcommand\B{\mathcal{B}}

\newcommand\C{\mathcal{C}}
\newcommand\D{\mathcal{D}}
\newcommand\E{\mathcal{E}}

\newcommand\F{\mathcal{F}}

\renewcommand\O{\mathcal{O}}

\renewcommand\S{\mathcal{S}}

\newcommand{\To}{\rightrightarrows}
\newcommand{\inv}[1]{#1^{-1}}

\newcommand{\type}{\tau}
\newcommand{\plan}{\zeta}

\newcommand{\mpty}{{\mathit{E}}} % \empty is a LaTeX macro that is
\newcommand{\Empty}{\textrm{`Empty'}}
\newcommand{\full}{{\mathit{F}}}
\newcommand{\Full}{\textrm{`Full'}}

\newcommand{\half}{{\mathit{H}}}

\newcommand{\A}{\mathsf{A}}
\renewcommand{\a}{\mathsf{a}}
\renewcommand{\d}{\mathsf{d}}
\newcommand{\w}{\mathsf{w}}
\newcommand{\e}{\mathsf{e}}
\newcommand{\f}{\mathsf{f}}
\newcommand{\h}{\mathsf{h}}

\newcommand{\setbrac}[1]{ \left\{ #1 \right\} }

\begin{document}

\begin{abstract}
People reason heuristically in situations resembling
inferential puzzles such as Bertrand's box paradox and the
Monty Hall problem. The practical significance of that fact
for economic decision making is uncertain because a departure
from sound reasoning may, but does not necessarily, result in
a ``cognitively biased'' outcome different from what sound
reasoning would have produced.  Criteria are derived here,
applicable to both experimental and non-experimental
situations, for heuristic reasoning in an inferential-puzzle
situations to result, or not to result, in cognitively
bias. In some situations, neither of these criteria is
satisfied, and whether or not agents' posterior probability
assessments or choices are cognitively biased cannot be
determined.
\end{abstract}

\maketitle

\Section{1}{Introduction}

People use heuristic reasoning in decision situations, and thus
potentially make ``cognitively biased'' decisions that deviate from
what they would have done if they had reasoned soundly. This article
concerns conditions under which a particular type of heuristic
Bayesian inference will, or will not, deviate from sound inference in
a situation, and provides examples of plans (that is, patterns of
evidence-based choices) that result from sound inference in some
situations, and from heuristic inference in others, while being
demonstrably inconsistent with the other sort of reasoning. This
explicit concern with demonstrability (with identifiability, in
statistical or econometric parlance), rather than with the simple
occurrence or non occurence of cognitive bias, may distinguish the
present research with respect to behavioral-economics research on the
whole.

The paradigmatic situation to be studied is the ``box
paradox'' formulated by
\citet[p.\ 2]{Bertrand-1889}. \citet{Gardner-1959} and others
have subsequently formulated isomorphic
problems. \citet{BarhillelFalk-1982} recognized and elucidated
the significance of those problems for cognitive psychology.
\citet{ShimojoIchikawa-1989} conducted a pioneering
cognitive-psychology experiment to understand better the logic
of the heuristic reasoning by which people analyze such
situations. \citet{Selvin-1972a}, inspired by an eponymous
television producer's adaptation of such a situation for
entertainment, formulated the ``Monty Hall
problem''.\footnote{Mark Feldman has mentioned to the authors
  that the Monty Hall problem is isomorphic to the situation
  of ``restricted choice'' in the game of bridge.} The
distinctive feature of this problem is that a person is
required to make a utility-maximizing choice among a set of
alternate gambles, rather than to express a numerical
probability judgment. That is, it is a ``behavioralistic'' (in
the sense of \citet{Savage-1972}) version of the box
paradox. \citet{GranbergBrown-1995}, followed by
\citet{Friedman-1998} and others, have used that problem as
the basis for an experimental protocol.

The experiments that have just been mentioned, are designed to
show a detectable, observable outcome from which an
unobservable cause can be inferred. The outcome is an
incorrect probability assessment or a biased decision, and the
cause is the subject's use of heuristic reasoning rather than
of sound reasoning. The import of the experiments is that,
even though the outcomes of heuristic reasoning are typically
not detectable by casual observation of non-experimental
situations (but, rather, require an insightfully designed
experimental protocol to become apparent), heuristic reasoning
is presumably endemic in everyday situations.

Heuristic reasoning is called `heuristic' for a reason: that
in some, albeit not all, cases that it is employed, it
providentially leads to correct or approximately correct
conclusions. Thus, that people endemically employ heuristic
inductive logic does not necessarily imply that faulty
posterior-probability assessments or misguided choices are
endemic.\footnote{Contemplating heuristic reasoning broadly,
  some researchers (such as \citet{Simon-1955}) have been
  inclined to believe that such cognitive bias is endemic in
  fact, and that experimental situations are exceptional only
  in point of the bias being demonstrable. Others (such as
  \citet{Friedman-1953}) have leaned toward the view that
  providential outcomes are normal, and that experimental
  situations are exceptional because cognitive bias occurs at
  all.} The program of this article is to examine, in the
specific context of situations resembling the box paradox and
the Monty Hall problem, what are the characteristics of
situations in which outcomes (posterior-probability
assessments or choices) are, or are not, informative about
whether cognitive bias has occurred. That is, the goal is to
distinguish among three types of situation.
\begin{itemize}
\item[\emph{Type 1}]{In some situations, including
  experiments, some outcomes may be observed that demonstrably
  reflect heuristic reasoning and are inconsistent with sound
  reasoning. That is, persons (or \emph{agents\/}) making
  those choices exhibit cognitive bias.}
\item[\emph{Type 2}]{There may also be situations in which
  heuristic reasoning will lead demonstrably to the same outcome
  as sound reasoning would have produced, given identical
  probability beliefs regarding potentially observable
  events. That is, even if the agent is reasoning
  heuristically, no cognitive bias will result from it.}
\item[\emph{Type 3}]{Finally, there may be situations in which
  some observable outcome can be imputed to heuristic
  reasoning by making one set of assumptions about the agent's
  beliefs (and about utilities of available alternatives, if
  the outcome is a choice or decision), but different
  assumptions about the agent lead to the conclusion that the
  same outcome has resulted from sound reasoning. That is,
  given that outcome, although there is cognitive bias if the
  first set of assumptions is correct, the bias is not
  demonstrable because the alternate set of assumptions
  cannnot be ruled out.}
\end{itemize}

Sections \secref{2} and \secref{3} are devoted to formalizing
a broad class of situations resembling the box paradox and
Monty Hall problem, and to articulating what it means for
heuristic reasoning to be justified by sound reasoning in such
a situation. \uppercase\prp{1} (in \sec{4}) provides a criterion for
posterior beliefs reached by heuristic reasoning to be
justifiable by sound reasoning, if specific prior beliefs are
imputed to the agent. But the criterion does not rule out the
possibility that those beliefs are cognitively biased outcomes
of different prior beliefs. That is, a situation that meets
the criterion might be of either type 2 or type 3. The
trichotomy of situations is studied further in sections
\secref{5} and \secref{6}. \uppercase\prp{2} (in \sec{6}) provides
conditions that are sufficient (and, under an auxilliary
assumption, necessary) for a situation to be of one or another
of the three types. \uppercase\sec{7} concerns the
``behavioralistic'' framework, in which outcomes of reasoning
are taken to be choices rather than posterior-probability
assessments. This framework invokes more parsimonious
assumptions about what is observable to a researcher than the
``verbalistic'' framework of sections \secref{3}--\secref{6}
makes. Not surprisingly, it becomes more difficult to infer
from outcomes how an agent has reasoned. Nonetheless,
\xmpl{4} exhibits a pattern of choices that can only arise as
an outcome of sound reasoning, while \xmpl{6} exhibits a
pattern that can only result from heuristic reasoning (and
thus is cognitively biased).

\Section{2}{An example of heuristic inference}

Here we formulate, and analyze in ad hoc terms, an example of the sort
of heuristic inference that is the subject of this
article.\footnote{The example is formulated to avoid some features of
  the Monty Hall problem that \citet{GranbergBrown-1995} and
  \citet{Friedman-1998} have identified as being related to other
  biases---involving revisions of choices and breaking of indifference
  among alternatives---to which some subjects' performance might be
  imputed. The relationship between the example and
  the Monty Hall problem will be examined in \sec{7.3}.} It
illustrates the type of bias that is routinely observed in the
performance of subjects in cognitive-science
experiments.\footnote{Early studies, such as those of
  \citet{GranbergBrown-1995} and \citet{Friedman-1998} established
  that about 90\% of subjects initially give biased responses that
  would be derived from the heuristic analysis to be specified below,
  and that roughly 50\% of subjects persist in giving those responses
  after many repetitions of the problem. Subsequent studies, such as
  the one by \citet{KlugerFriedman-2010} and those that they cite,
  establish that some experimental treatments can reduce the incidence
  of cognitively biased responses, but not dramatically so.}

\Subsection{2.1}{The broken-fuel-gauge (BFG) problem}

Your car has a broken fuel gauge. It always shows either \Full\ or
\Empty. When the tank is more than $70\%$ full, the gauge always shows
\Full. When the tank is less than $30\%$ full, the gauge always
 shows \Empty. In between, the gauge might be in either state.

You have been on vacation---away from your car---for a month. You no
longer recall how far you drove after last having filled the
tank. Before reading the gauge, your beliefs about the amount of fuel
in the tank correspond to a uniform distribution.

When you look, the gauge shows \Empty.  What is now your degree of
belief that the tank is at most $30\%$ full? In the notation of
probability theory, what is $P[\text{Tank is\ }\! \le 30\%
  \text{\ full} \mid \text{\Empty}]$?

\Subsection{2.2}{Heuristic analysis}

Let $F_x$ denote the event that the tank is at most $x\%$ full. Then $P(F_x)
= x/100$.

The \emph{heuristic analysis} of the BFG problem is based on the
assumption that the gauge showing \Empty\ corresponds to $F_{70}$. Of
course, there are other assumptions that an agent might conceivably
substitute for the more complex and subtle assumtion that sound
reasoning would require, but this particular assumption is one that
succeeds in accounting for way that experimental subjects tend to
respond to such situations.

To an agent who reasons heuristically, then, a particular
configuration of the gauge denotes the set of states of nature
in which the gauge can possibly be in that configuration. If
you reason heuristically, then, when asked what is $P[F_{30}
  \vert \text{\Empty}]$, you interpret that conditional
probability as being $P[F_{30} \vert F_{70}]$. By Bayes's
rule,
\display{1}{P[F_{30} \vert F_{70}] = \frac{P(F_{30} \cap
    F_{70})}{P(F_{70})} = \frac{P(F_{30})}{P(F_{70})} = \frac{3}{7}}

\Subsection{2.3}{Sound analysis}

A conceptually correct, or \emph{sound,} analysis of the BFG problem
proceeds according to the logic articulated by
\citet{Harsanyi-1967}. This analysis emphasizes that your having
observed the tank to show \Empty\ is a fact about you, rather than
being \emph{per se} a fact about the tank or its contents. 
Whether the gauge shows \Empty\ or \Full\ determines your
\emph{type.}\footnote{When this example is formalized below, a third
  type---corresponding to you not having yet observed the gauge (and
  thus holding your prior beliefs)---will be added. That change will
  not affect the present calculation.}

Your type is random, from an \emph{ex ante} point of
view. This randomness is modeled as a type-valued function
$\tau \from \Phi \to \{ \Empty, \Full \}$, where $\Phi$ is the
set of states of the world.  The gauge showing \Empty\ (and
you observing that fact) corresponds to the event that the
state of the world is in $\inv \type(\Empty)$. Thus, in
contrast to the heuristic analysis, $P[F_{30} \vert
  \text{\Empty}]$ means $P[F_{30} \vert \inv \type(\Empty)]$.

\display{2}{P[F_{30} \vert \inv \type(\Empty)] = \frac{P(F_{30} \cap
  \inv \type(\Empty))}{P(\inv \type(\Empty))}
= \frac{P(F_{30})}{P(\inv \type(\Empty))}}

Note that $\inv \type(\Empty) = F_{30} \cup (\inv
\type(\Empty) \cap (F_{70} \setminus F_{30}))$. Implicit in
the specification that ``The gauge might show either
\Empty\ or \Full\ in event $F_{70}$,'' is the idea that
$P(\inv \type(\Empty) \cap (F_{70} \setminus F_{30})) <
P(F_{70} \setminus F_{30})$.\footnote{This idea is formalized
  for heuristic reasoning in the third clause of condition
  \eqn{7} below. By condition \eqn{20}, the idea extends to
  sound reasoning also.} Therefore,
\display{3}{P[F_{30} \vert \inv \type(\Empty)] > \frac{P(F_{30})}{P(F_{70})} = \frac{3}{7}}

In conclusion, comparing \eqn{1} and \eqn{3} shows that the sound
analysis yields a higher answer than the heuristic analysis does to
the question about your posterior belief that the tank is truly near
empty after having observed \Empty.

\Section{3}{Models of evidence and of beliefs}

Two types of structures will be defined in this section, and how they
apply to the BFG problem will be explained. A \emph{model
  of evidence} formalizes heuristic Bayesian inference. A \emph{model
  of beliefs} formalizes sound Bayesian inference.  According to a
model of beliefs, the agent reasons introspectively about the grounds
for his/her beliefs. That is, the agent asks, what determines the
relationship of its own cognitive state to the objective facts about
the world? An agent whose reasoning is represented by a model of
evidence, is not introspecting. Rather, the agent is thinking solely
in terms of objective events. Within each framework, the agent revises
beliefs (that is, subjective probabilities) according to Bayes's rule.
The question to be addressed is under what conditions a model of
evidence reflects---that is, is justified by---a model of beliefs.

\Subsection{3.1}{Model of evidence} A model of evidence is a structure,
$(\Omega, \O, P, \E)$, where
\begin{enum}
\meti{4}
{$\Omega$ comprises the \emph{states of nature}}
\meti{5}
{$\O$ is a $\sigma\/$-field of \emph{objective events} on $\Omega$}
\meti{6}
{$P \from \O \to [0,1]$ is a countably additive probability measure}
\meti{7}
{$\E \subseteq \O$ comprises the \emph{evidential events.}}
\bi
\item
$\E$ is countable
\item
$B \in \E \implies P(B) > 0$
\item
$[B \in \E \mand C \in \E \mand C \not \subseteq B] \implies
  P(C \setminus B) > 0$
\item
$\Omega \in \E$
\item
$\bigcup (\E \setminus \setbrac{\Omega}) = \E$
\ei
\end{enum}

Clearly an agent requires no evidence to be certain
that $\omega \in \Omega$. It will be convenient to have a
notation for the non-trivial evidential events, that is, for
those in $\E \setminus \setbrac{\Omega}$. Define
\display{8}{\E' = \E \setminus \setbrac{\Omega}}

The assumptions made in condition \eqn{7} reflect the focus of
this article. Notably the assumptions that $P(B) > 0$ and that
$P(C \setminus B) > 0$ simplify the analysis of the specific
cognitive bias studied here, to which subtle questions that
arise concerning conditioning on events of prior probability
zero have no apparent relevance. That is, although questions
regarding how to extend conditional probability to
conditioning events of prior probability zero are crucial for
some issues in game theory, they are arcane in the context of
belief revision and choice by a single agent.

It is assumed that $\E$ is countable because, otherwise, that $P(B) > 0$
for every $B \in \E$ would be impossible.\footnote{It would be possible to
define a more general structure that would not require $\E$ to be
countable, analogously to the way that full-support probability
distributions on continuously distributed random variables are defined
in probability theory, but to do so would greatly complicate the
mathematical arguments to be made here without making a corresponding
conceptual gain.}

The assumption that $\Omega \in \E$ is a convention that will
play a role in defining what it means for a model of beliefs
to justify a model of evidence. $\E'$ is actually the set of
entities that formalizes the intuitive notion of a non-trivial
evidential event. In principle, there might be some state of
the world for which no corroborating evidence could possibly
be found. That is, conceivably $\bigcup \E' \neq \Omega$. A
condition, \emph{balancedness,} will be defined in \sec{4.1},
that will fail if $P(\bigcup \E') < 1$. \uppercase\prp{1} will assert
that balancedness of a model of evidence is a necessary and
sufficient condition for there to be some model of beliefs
that justifies it. Thus it is known that if, the definition of
a model of evidence were relaxed to permit that $P(\bigcup
\E') < 1$, then such model would represent a situation of type
1 (in the taxonomy of the introduction). But, rather than
complicate the exposition of \prp{1} and other results by
explicit consideration of that possibility of un-corroborable
states of the world (and to avoid arcane complications of
dealing with probability-zero events), it has been stipulated
that $\bigcup \E' = \Omega$.

As usual in Bayesian decision theory, the probability space, $(\Omega,
\O, P)$ models an agent's prior beliefs. The events in $\E'$
model observations that the agent might make, on the
basis of which evidence the agent would form posterior beliefs. Those
beliefs are formed by conditionalization, where conditional
probability,\\ $P \from \O \times \E \to [0,1]$ is defined as
usual:\footnote{Throughout this article, $A$ should be interpreted to
  range over all of $\O$, and $B$ to range only over $\E$, absent a
  statement to the contrary.}
\display{9}{P[A | B] = \frac{P(A \cap B)}{P(B)}}

Let's see how the heuristic analysis of the BFG problem is formalized
as a model of evidence. The description of the problem in \sec{2.2} is
made more simple here, by assuming that there are just three states of
nature, $\Omega = \{ e,h,f \}$. State $e$ represents the situation
that the fuel tank is nearly empty ($0 \le x < 30$ in the setting of
\sec{2.2}); $h$ represents the situation that the fuel tank is half
full ($30 \le x < 70$); and $f$ represents the situation that the
fuel tank is nearly full ($70 \le x \le 100$).

\exmpl{1}{Define $(\Omega, \O, P, \E)$ as follows. $\Omega = \{ e,h,f
  \}$ and $\O = 2^\Omega$. By additivity, $P$ is defined by the
  probabilities of singleton events in $\O$. Specify that $P( \setbrac{e} )
  = P( \setbrac{f} ) = 0.3$ and that $P( \setbrac{h} ) = 0.4$.  There are two
  non-trivial evidential events: $\mpty = \{ e, h \}$ and $\full = \{
  h,f \}$. $\E = \{ \Omega, \mpty, \full \}$.}

Note that, corresponding to \eqn{1} in \sec{2.2}, $P[\setbrac{e} | \mpty] =
3/7$.

\Subsection{3.2}{Model of beliefs}

\citet{Harsanyi-1967} introduced a structure that he called a beliefs
space, which consists of a probability space, $(\Phi, \B, Q)$ and a
\emph{type mapping,} $\type \from \Phi \to T$, where $T$ is an abstract
set. The elements of $\Phi$ are \emph{states of the world} and the
elements of $T$ are \emph{types} of the agent. The types in Harsanyi's
structure, \emph{per se,} are nothing but arbitrary labels. What is
meaningful are the inverse images, $\inv \type(t)$, of the
types. These \emph{information sets} partition the states of the
world. In the event that an agent is of type $t$, then the agent's
posterior belief regarding an event, $C$, is $Q[C|\inv\type(t)]$,
where $Q \from \B \times \inv\type(T) \to [0,1]$ is defined
analogously to \eqn{9}.

A model of beliefs, $(\Phi, \B, Q, \Omega, \E, \type)$, is a slight variant of
a beliefs space.
\begin{enum}
\meti{10}
{$\Omega$ comprises the \emph{states of nature.} $\E \subset 2^\Omega
  \setminus \setbrac{\emptyset}$ is countable. $\Omega = \bigcup \E' \in \E$.}
\meti{11}
{$\Phi$ comprises the
  \emph{states of the world.} $\B$ is a $\sigma\/$-field on $\Phi$ and
  $Q \from \B \to [0,1]$ is a countably additive probability measure.}
\meti{12}
{The \emph{type function,} $\type \from \Phi \to \E'$, maps
  $\Phi$ \emph{onto} $\E'$. $\Omega$ is called the \emph{prior-beliefs
    type,} and elements of $\E'$ are called \emph{posterior-beliefs
    types.}\footnote{This nomenclature corresponds to standard
    terminology in decision theory regarding a single agent. In game
    theory, probability assessments conditioned on a player's type are
    generally called \emph{interim beliefs.}}}
\meti{13} {$\inv\type(B) \in \B$ and $0 < Q(\inv\type(B)) < 1$}
\end{enum}
It will be convenient to extend $\inv\type$, the inverse
correspondence of $\type$, to a correspondence, $\beta \from \E \to
\B$.
\display{14}{\phi \in \beta(B) \iff [B = \Omega \text{\ or\ }
    \type(\phi) = B]}

It is a trivial formal change of Harsanyi's framework to specify
that the agent's types are sets of states of nature rather than being
arbitrary labels, and to introduce a new type that is not realized in
any state of the world. Nonetheless, this modification enables a
model of evidence and a beliefs space to be compared explicitly as
representations of Bayesian inference. Before framing a systematic
comparison in the next section, let's see how a model of belief
contrasts with a model of evidence as a representation of inference in
the BFG problem. The intuition behind the formalization of the BFG
problem as a model of beliefs is as follows. $\Omega$, $P$, and $\E$
are as in \xmpl{1}. Type $\Omega$ represents the agent's prior
beliefs, while types $\mpty$ and $\full$ represent posterior beliefs
after having observed the gauge to show \Empty\ and
\Full\ respectively. Conditionally on
the state of nature being $e$ or $f$, if the agent's type is not
$\Omega$, then it must be $\mpty$ or $\full$ respectively. However, if
the state of nature is $h$ and the agent's type is not $\Omega$, then
the type may be either $\mpty$ or $\full$. Assume that, conditionally on the
state of nature being $h$ and the agent's type not being $\Omega$, the
other two types are equally probable.

\exmpl{2}{Let $(\Omega, \B, P, \E)$ be as in \xmpl{1}. $\Phi = \Omega
  \times \E' = \{ e,h,f \} \times \{ \mpty, \full \}$. $\B =
  2^\Phi$. $Q$ is specified according to the following table. Each
  cell of the table is a probability. The column
  labeled `$\Omega$' shows marginal probabilities of $Q$ on
  $\Omega$. The cell in the row labeled `$\omega\/$' and the column
  labeled `$B\/$', for $B \in \E'$, shows the probability of the
  corresponding state of the world, $(\omega,B)$.
\display{15}{ \begin{tabular}[c]{|c||c|c|c|}
\hline
$Q(\omega,B)$ & $\Omega$ & $\mpty$ & $\full$\\

\hline \hline

$e$ & .3 & .3 & 0\\

\hline

$h$ & .4 & .2 & .2\\

\hline

$f$ & .3 & 0 & .3\\

\hline

\end{tabular} }
\hide{
$Q( \setbrac{e} \times \{ \mpty \} ) = Q( \setbrac{f} \times \{
  \full \} ) = 0.3$. $Q( \setbrac{e} \times \{ \full \} ) = Q( \setbrac{f}
  \times \{ \mpty \} ) = 0$. $Q( \setbrac{h} \times \{ \mpty \} ) = Q( \{
  h \} \times \{ \full \} ) = 0.2 = P( \setbrac{h} )/2$.
}
The type function is defined by $\type(\omega,B) = B$ for all $\omega
\in \Omega$ and $B \in \E'$. Note that $Q(A \times \Omega) = P(A)$ for
all $A \in \O$, where $P$ is the probability measure constructed in
\xmpl{1}.}

If $A$ is a set of states of nature, then the event that the state of
nature is in $A$ is $A \times \E' \in \B$. Thus, as explained above,
the posterior probability held by an agent of type $B \in \E'$ that the
state of nature is in $A$ is $Q[A \times \E' | \beta(B)]$. In
particular, taking $A = \{e \}$ and $B = \mpty$, the agent's posterior
probability that the state of nature is $e$ is $3/5$. Since it has
been shown that $P[\setbrac{e} | \mpty] = 3/7$ in \xmpl{1}, the formal
representations of the BFG problem via a model of evidence and a model
of belief reproduce the overall conclusion, inequality \eqn{3}, of \sec{2.2}.

\Subsection{3.3}{Reflection/justification}

In the context of comparing examples \rsltref{1}{xmpl} and
\rsltref{2}{xmpl}, it was just suggested that $A \times \E'$ is the
event in $\B$ that is associated with a set, $A \in \O$, of states of
nature. In fact, this association defines an isomorphism, $A \mapsto A
\times \E'$, of $\O$ with a sub $\sigma\/$-field of $\B$ in the
example. Capitalizing on this idea, the relationship between a model
of evidence and a model of beliefs that was implicitly defined in that
discussion can be stated in an explicit and general way.

In order to make this generalization, an isomorphism of probability
spaces must be defined in a slightly more permissive way than the
obvious one. In the preceding paragraph, `isomorophism' was used in
the obvious way, to denote a mapping that preserves Boolean
relationships among sets. However, in general---when the situation
does not have the convenient feature that the domain of the
isomorphism is a Cartesian factor of its range---such an exact
relationship may not hold. Rather, if $(\Xi, \C, R)$ and $(\Psi, \D,
S)$ are probability spaces, then define $\alpha \from \C \to \D$ to be
a \emph{measure isomorphism} from $(\Xi, \C, R)$ to $(\Psi, \D, S)$ iff
the following conditions hold.\footnote{Definition \eqn{16} is tantamount to
  the definition of a measure isomorphism given by
  \citet[p.\ 202]{Sikorski-1969}.}
\display{16}{\begin{aligned}
&\text{For every $C \in \C$,} \enspace S(\alpha(C)) = R(C) \\
&\text{For every countable $\F
  \subseteq \C$,} \enspace S \left(\ \alpha \left( \bigcup \F \right)
  \triangle \bigcup \{ \alpha(B) \mid B \in \F \} \right) = 0\\
&\text{For every $D \in \D$, there exists $C \in \C$ such
  that} \enspace  Q(D \triangle \alpha(C)) = 0
\end{aligned}}
Throughout the remainder of this
article, \emph{isomorphism} refers to a measure isomorphism. When the
full specifications of $(\Xi, \C, R)$ and $(\Psi, \D, S)$ are clear
from context, $\alpha$ will be called an isomorphism from $\C$ to $D$.

Let $\Omega \in \E \subseteq 2^\Omega \setminus \{ \emptyset \}$.
A model of beliefs, $(\Phi, \B, Q, \Omega, \E, \type)$, \emph{conforms} to a
model of evidence, $(\Omega, \O, P, \E)$, via $\alpha \from \O \to \B$ iff
\begin{enum}
\meti{17}
{For some sub $\sigma\/$-field, $\C$, of $\B$,
$\alpha \from \O \to \B$ is a measure isomorphism from $(\Omega, \O,
  P)$ to $(\Phi, \C, Q \upharpoonright \C), and$}
\meti{18}
{For all $B \in \E$, $\beta(B) \subseteq \alpha(B)$}
\end{enum}

A model of evidence, $(\Omega, \O, P, \E)$, \emph{reflects} a model of
beliefs, $(\Phi, \B, Q, \Omega, \E, \type)$, (equivalently, the model of
beliefs \emph{justifies} the model of evidence) via $\alpha \from \O
\to \B$ under the following conditions.\footnote{Subsequently, unless
  specific information about $\alpha$ is important to a discussion,
  ``via $\alpha$'' will sometimes not stated explicitly, although 
  conditions stated in terms of $\alpha$ may be invoked.} 
\begin{enum}
\meti{19}
{$(\Phi, \B, Q, \Omega, \E, \type)$, conforms to $(\Omega, \O, P, \E)$, via
  $\alpha$, and}
\meti{20}
{For all $A \in \O$ and $B \in \E$, $Q[\alpha(A) \vert \beta(B)] = P[A
    \vert B]$}
\end{enum}

Note that, when this relationship holds, there is a clear reason to
view the agent's types as being evidential events rather than mere
abstract labels. The agent's type is the most specific objective event
(not only the most specific evidential event) that the agent believes
to obtain almost surely. That is, for any $A \in \O$ and $B \in \E$,
$Q(\beta(B) \setminus \alpha(B)) = 0$ and, if $P(A) < P(B)$, then
$Q(\beta(B) \setminus \alpha(A)) > 0$.

Condition \eqn{20} is central to this article, because it formalizes
what it means for cognitive bias \emph{not} to occur. An agent is
envisioned to have authentic probability beliefs, either fully
articulated or inchoate, that are represented by a model of beliefs,
$(\Phi, \B, Q, \Omega, \E, \type)$. With respect to the events in some sub
$\sigma\/$-field of $\B$, at least, the agent's beliefs are envisioned
to be fully articulated. Specifically it is envisioned that there is a
model of evidence, $(\Omega, \O, P, \E)$, such that $(\Phi, \B, Q, \Omega, \E,
\type)$ conforms to $(\Omega, \O, P, \E)$ via some measure
isomorphism, $\alpha$, and that the agent's authentic probability
beliefs about events in the image of $\O$ under $\alpha$ are fully
articulated. That is, condition \eqn{19} is satisfied. What determines
whether or not $(\Omega, \O, P, \E)$ reflects $(\Phi, \B, Q, \Omega, \E,
\type)$, is condition \eqn{20}. If \eqn{20} is not satisfied, then
$(\Omega, \O, P, \E)$ does not reflect $(\Phi, \B, Q, \Omega, \E,
\type)$. Intuitively that is the case in which, if the agent reasons
heuristically according to the model of evidence, $(\Omega, \O, P,
\E)$, then the agent exhibits cognitive bias relative to sound
inference from authentic beliefs (that is, relative to inference based
soundly on the model of beliefs, $(\Phi, \B, Q, \Omega, \E, \type)\,\strut$).

\claim{3}{The model of evidence in \xmpl{1} does not reflect any model
  of beliefs. The model of beliefs in \xmpl{2} does not justify any
  model of evidence.}
\begin{proof}
Suppose that $\alpha \from \O \to \B$ satisfies conditions \eqn{17}
and \eqn{19}, and that either $(\Omega, \O, P, \E)$ is the model of
evidence in \xmpl{1} or else $(\Phi, \B, Q, \Omega, \E, \type)$ is the model of
beliefs in \xmpl{2}. In either case, $P(\mpty) = P(\full) = 7/10$ and
$P( \setbrac{h} ) = 4/10$. so $P[ \setbrac{h} | \mpty] = P[ \setbrac{h} | \full] =
4/7$.  Since $(\Phi, \B, Q, \Omega, \E, \type)$ is a model of beliefs, $0 <
\min \{ \beta(\mpty),\beta(\full) \}$ and $\{ \beta(\mpty),
\beta(\full) \}$ partitions $\Phi$. Therefore $\{ \alpha( \setbrac{h} )
\cap \beta(\mpty), \alpha( \setbrac{h} ) \cap\beta(\full) \}$ partitions
$\alpha( \setbrac{h} )$.

A contradiction will be derived from the assumption that condition
\eqn{20} is also satisfied (that is, that $(\Omega, \O,
P, \E)$ reflects $(\Phi, \B, Q, \Omega, \E, \type)$).  Since $P[ \setbrac{h}
  |\full] = 4/7$, condition \eqn{20} requires that $Q(\alpha( \setbrac{h}
) \cap \beta(\full)) = 4Q(\full)/7 > 0$. Therefore $Q(\alpha( \setbrac{h}
) \cap \beta(\mpty)) < Q(\alpha( \setbrac{h} ))$. Then, by \eqn{19},
\display{21}{\begin{gathered}
Q[\alpha( \setbrac{h} ) | \beta(\mpty)] = Q(\alpha( \setbrac{h} ) \cap
\beta(\mpty))/Q(\beta(\mpty))\\ < Q(\alpha( \setbrac{h} ))/Q(\beta(\mpty)) =
P( \setbrac{h} )/P(\mpty) = P[ \setbrac{h} | P(\mpty) ]
\end{gathered}}
This contradicts \eqn{20}.
\end{proof}

A fact about conformity, to be used later in the proof of \prp{2},
is stated and proved now.

\lema{1}{For every model of evidence, there is a conforming model of
  beliefs.}

\begin{proof}
Consider a model of evidence, $\Omega, \O, P, \E)$. Let $\langle B_s
\rangle_{s \in S}$ enumerate $\E'$. ($S = \{ 1,\dotsc, n \}$ if $\E'$
has $n$ elements, and $S = \setbrac{1,2,3\dots}$ if $\E'$ is
infinite.) Let $t = \sum_{s \in S} 2^{-s}$. Define $\Phi = \Omega
\times \E'$ and $\B = \Sigma(\O \times 2^{\E'})$.\footnote{If $\C
  \subseteq 2^\Phi$, then $\Sigma(C)$ is the smallest $\sigma\/$-field
  containing $\C$.} Begin to  define $Q$ by $Q(A \times B_s) = 2^{-s}
P(A)/t$. This definition extends by countable additivity to $\Sigma(\O
\times 2^{\E'})$.\footnote{Specifically, this extension is a measure
  by Caratheodory's theorem. (Cf.\ \citet[theorem
    10.23]{AliprantisBorder-2006}).} Define $\mu \from \Omega \to S$
by $\mu(\omega) = \min \setbrac{ s \mid \omega \in B(s)}$, and define 
$\type \from \Phi \to \E'$ by
\display{22}{\type(\omega, B) = \begin{cases}
B & \text{if\ } \omega \in B\\
B_{\mu(\omega)} & \text{if\ } \omega \notin B
\end{cases}}
Define $\alpha \from \O \to \B$ by $\alpha(A) = A \times \E'$.
It is routinely verified that $(\Phi, \B, Q, \Omega, \E, \type)$ is a model
of beliefs that conforms to $(\Omega, \O, P, \E)$ via $\alpha$.
\end{proof}

\Section{4}{When does some model of beliefs justify a model of evidence?}

In this section, a necessary and sufficient condition will be derived
for a model of evidence to reflect some model of beliefs.
To set the stage, let us point out a feature of the BFG problem that seems to
be conducive for cognitive bias to occur. Consider the model of
beliefs presented in \xmpl{2}. There are only 2 posterior-beliefs types:
$\mpty = \{ e,h \}$ and $\full = \{ h,f \}$. The image of $\setbrac{h}$
under $\alpha$ is split between $\beta(\mpty)$ and $\beta(\full)$. 
Its probability mass is correspondingly split in the model of beliefs.

In contrast, $\alpha( \setbrac{e} ) \subseteq \beta(\mpty) \mand \alpha(
\setbrac{f} ) \subseteq \beta(\full)$. The probability mass of these
states of nature therefore is not split.

Reflection is impossible because probability mass of some states of
nature, but not others, must be split. This situation results from a
particular state of nature being in both $\mpty$ and $\full$, while
others are only in one of them. A state of nature that belongs
to more
evidential events than others do, is
under weighed relative to those others by $Q$ within the image under $\beta$ of each of
the evidential events to which it belongs.

Clearly this problem of under weighting cannot occur if $\E'$ is a
partition of $\Omega$. The following example shows that, to avoid
under weighting, it is not necessary for $\E'$ to be a partition, or
even for there to exist a partition of $\Omega$ by elements of
$\E'$. A condition that the example does satisfy, and that will be
generalized below, is that each state of nature belongs to the same
number (3, in the example) of evidential events.

\exmpl{3}{Define $(\Omega, \B, P, \E)$ by setting $\Omega = \{ 0,1,2
  \}$, $\B = 2^\Omega$, and $P(\{ \omega \}) = 1/3$, and by defining
  $\E'$ to be the set of two-element subsets of $2^\Omega$. For each
  $\omega \in \Omega$, define $\omega' \equiv \omega+1 \pmod 3$ and
  $\omega'' \equiv \omega-1 \pmod 3$, and define $E_\omega = \{
  \omega, \omega' \}$.\footnote{The states of nature and the
    evidential events in \protect\xmpl{1} can be embedded in this
    structure by assigning $e \mapsto 0$, $h \mapsto 1$, $f \mapsto
    2$, $\mpty \mapsto B_0$, and $\full \mapsto B_1$.}  Define $\Psi =
  \{ 0,\dotsc,5 \}$, $\C = 2^\Psi$, and define $R( \{ \psi\} ) = 1/6$
  for each $\psi$. Let $(\Phi, \B, Q)$ be the product of $(\Omega, \O,
  P)$ and $(\Psi, \C, R)$. Using the unique representation of $\psi =
  2j+k$ (with $0 \le j \le 2$, $0 \le k \le 1$), define $\type$ by
\display{23}{\begin{gathered}
\type(\omega, \psi) = E_i \text{, where\ } i =
\begin{cases}
\omega, \text{\ if\ } k = 0\\
\omega'', \text{\ if\ } k = 1
\end{cases} \end{gathered}}
}

There is no partition of $\Omega$ by elements of $\E'$. Nonetheless,
defining $\alpha(A) = A \times \Psi$, $(\Phi, \B, Q, \Omega, \E, \type)$
justifies $(\Omega, \O, P, \E)$.

\Subsection{4.1}{Balancedness defined}

Define $(\Omega, \O, P, \E)$ to be \emph{balanced} iff, for
some $\theta$,
\display{24}{\theta \from \E' \to \left( 0,1 \right] \quad \mand \quad P \left( \left\{
\omega \mid \sum_{\omega \in B \in \E'} \theta(B) = 1 \right\} \right) = 1}
Call $\theta$ a \emph{balancing function} (for $P$ and $\E$).

Note that, with $\chi_B \from \Omega \to \setbrac{0,1}$ being the indicator
function of $B$, \eqn{24} is equivalent to
\display{25}{{\theta \from \E' \to \left( 0,1 \right] \quad \mand \quad  P \left( \left\{
\omega \mid \sum_{B \in \E'} \theta(B) \chi_B(\omega) = 1 \right\} \right) = 1}
}

By setting $\theta(B_0) = \theta(B_1) =
\theta(B_2) = 1/2$, it is seen that the model of evidence in \xmpl{3}
is balanced. In contrast, for the model of evidence in \xmpl{1}, if
$\theta \from \E \to (0,1)$, then $\sum_{h \in B} \theta(B) -
\sum_{e \in B} \theta(B) = \theta(\full) > 0$, so either $\sum_{h \in
  B} \theta(B) > 1$ or $\sum_{e \in B} \theta(B) < 1$ and therefore
\eqn{24} cannot hold. In each of these two examples, then, the model of
evidence being balanced is equivalent to it reflecting some model
of beliefs.

\Subsection{4.2}{Balancedness and the justifiability of a model of
  evidence}

The following proposition follows immediately from the two lemmas that
are subsequently proved.

\propo{1}{A model of evidence is balanced if, and only if, it reflects
  some model of beliefs.}

\lema{2}{A model of evidence is balanced, if some model of belief
  justifies it.}

\begin{proof}
Suppose that $(\Phi, \B, Q, \Omega, \E, \type)$ justifies $(\Omega, \O, P,
\E)$. By conditions \eqn{12}, \eqn{13} and \eqn{14}, $\setbrac{
  \beta(B) \mid B \in \E'}$ is a measurable partition of $\Psi$. By
  \eqn{16}, \eqn{17} and \eqn{20}, $Q(\alpha(A) \cap \beta(B)) =
P[A|B] \, Q(\beta(B))$.

Define $\theta \from \E' \to \left( 0,1 \right]$ by
\display{26}{\theta(B) = \frac{Q(\beta(B))}{P(B)}}
Then
\display{27}{\begin{gathered}
P(A) = Q(\alpha(A)) = \sum_{B \in \E'} Q(\alpha(A) \cap \beta(B)) =
\sum_{B \in \E'} P[A|B] \, Q(\beta(B))\\
=\sum_{B \in \E'} \int_A \frac{\chi_B}{P(B)} Q(\beta(B)) \, dP
= \int_A \sum_{B \in \E'} \theta(B)
\chi_B \, dP
\end{gathered}}

Given that \eqn{27} holds for all  $A \in \O$, condition \eqn{25} is
satisfied, so equation \eqn{26} defines a balancing function.
\end{proof}

\lema{3}{If a model of evidence is balanced, then some model
  of beliefs justifies it.}

\begin{proof}
Let $\theta$ be a balancing function for a model of evidence,
$(\Omega, \O, P, \E)$. A model of beliefs, $(\Phi, \B, Q, \Omega, \E, \type)$,
that justifies $(\Omega, \O, P, \E)$ is now constructed.

Let $\Psi = \left[ 0,1 \right)$ and let $\C$ and $R$ be the
  $\sigma\/$-field of Borel sets on $\Psi$ and the Lebesgue measure.
  Specify that $\Phi = \Omega \times \Psi$, $\B = \Sigma(\O \times
  \C)$, and $Q = P \times R$.\ Define an isomorphism, $\alpha \from \O
  \to \B$, by
\display{28}{\alpha(A) = A \times \Psi}

Let $\langle B_s \rangle_{s \in S}$ enumerate $\E'$. For $n \in \{ 0
\} \cup S$, define
\display{29}{g_0(\omega) = 0 \mand g_{n+1}(\omega) = g_n(\omega) +
  \theta(B_{n+1}) \chi_{B_{n+1}}(\omega)}

If $\langle x_s \rangle_{s \in S}$ is a sequence of numbers, then
define $\lim_{s \to \max S} x_s = x_{\max S}$ if $S$ is finite and
$\lim_{s \to \max S} x_s = \lim_{s \to \infty} x_s$ if $S$ is infinite.
Define $N = \{ \omega \mid \lim_{s \to \max S} g_s(\omega) \neq 1 \}$.
Since $\theta$ is a balancing function,
\display{30}{P(N)  = 0}
Define $\type \from \Phi \to \E'$ by
\display{31}{\type(\omega, \psi) = B_s \iff \begin{cases}
g_{s-1}(\omega) \le \psi < g_s(\omega) &\text{if\ } s \notin N\\
s = 1 &\text{if\ } s \in N
\end{cases}}
From this definition, it follows that
\display{32}{\begin{gathered}
 \{ (\omega, \psi) \mid \omega \in B_s \mand
  g_{s-1}(\omega) \le \psi < g_s(\omega) \} \subseteq 
\beta(B_s)\\ \subseteq
\{ (\omega, \psi) \mid \omega \in B_s \mand
  g_{s-1}(\omega) \le \psi < g_s(\omega) \} \cup N
\end{gathered}}
Then, from Fubini's theorem and \eqn{30} and \eqn{32}, it follows
that, for all $A \in \O$,
\display{33}{\begin{aligned}
Q(\alpha(A) \cap \beta(B_s)) &= 
\int_{A \cap B_s} \int_{g_{s-1}(\omega)}^{g_s(\omega)} 1 \, dR \, dP \\
&= \int_{A \cap B_s} g_s(\omega) - g_{s-1}(\omega) \, dP =
\theta(B_s) P(A \cap B_s)
\end{aligned}}

Conditions \eqn{30} and \eqn{32} and \eqn{33} imply that
\display{34}{Q(\beta(B_s)) = Q(\alpha(B_s) \cap \beta(B_s)) = \theta(B_s) P(B_s)}
so, for $A \in \O$ and $B \in \E$,
\display{35}{Q[\alpha(A) | \beta(B)] = \frac{Q(\alpha(A) \cap
    \beta(B))}{Q(B)} = \frac{\theta(B)P(A \cap B)}{\theta(B)P(B)} = P[A | B]}
That is, $(\Phi, \B, Q, \Omega, \E, \type)$ justifies $(\Omega, \O, P, \E)$.
\end{proof}

\Section{5}{Situations}

An undefined term, \emph{situation}, played an important role in the
introductory discussion of alternate views about the import of
psychological experiments for economics. In this section, building on
the framework introduced in the preceding section, a situation will be
formally defined.

In the introduction, it was mentioned that
\citet{ShimojoIchikawa-1989} used a situation isomorphic to Bertrand's
box paradox as the basis for an experiment to exhibit subjects'
cognitive bias. In that situation, there are three states of
nature. Shimojo and Ichikawa stipulated that, given the way that the
situation was described to subjects in the experimental protocol,
their prior beliefs would be that each state of nature has probability
$1/3$. (That is, those would be the subjects' probability assessments
after having received the description of the situation, but before
having observed evidence that would be presented in the course of the
experiment.) However, those researchers did not make any assumption
regarding a subject's beliefs about the correlation between the state
of nature and the evidence that would be observed. They did not need
to make any such assumption, because most subjects reported posterior
probability assessments that were inconsistent with \emph{any} model
of beliefs corresponding to the stipulated prior probability beliefs
about states of nature. That is, the outcomes of the experiment were
generated by a situation of type {1} according to the trichotomy
presented in the introduction.

The idea of a ``model of beliefs corresponding to the stipulated prior
probability beliefs about states of nature'' is formalized by the
definition of conformity. Shomojo and Ichikawa's assumptions about
subjects' beliefs regarding the states of nature can be represented as
a model of evidence. As has been discussed following the definition of
reflection in the previous section, their discussion of their
experiment presupposed that each subject had authentic probability
beliefs about the state of the world that could be represented as some
model of beliefs or other, but they did not pretend to know anything
about that model beyond the fact that it conformed to the stipulated
model of evidence. The general form of Shimojo and Ichikawa's way of
thinking about their experiment is captured by the following
definition. That is, a situation is a structure that formally
describes a researcher's assumptions regarding both the observable and
unobservable aspects of a what that researcher assumes to be a
subject's authentic prior-probability beliefs. Specifically, the
research assumes everything that is common to all of the models of
belief in the situation.

A \emph{situation} is a structure, $((\Omega, \O, P, \E), \S)$,
comprising a model of evidence and a non empty set, $\S$, of ordered pairs. Each
element of $S$ is of the form $((\Phi, \B, Q, \Omega, \E, \type), \alpha)$, where
$(\Phi, \B, Q, \Omega, \E, \type)$ is a model of beliefs that conforms to
$(\Omega, \O, P, \E)$ via $\alpha \from \O \to \B$. Where no confusion
will result from abuse of notation, `$\S\/$' will be used to name
the situation. Also, a statement such as ``Some model of beliefs in
$\S$ justifies $(\Omega, \O, P, \E)$'' should be understood as
``For some $((\Phi, \B, Q, \Omega, \E, \type), \alpha) \in \S$,
$(\Phi, \B, Q, \Omega, \E, \type)$ justifies $(\Omega, \O, P, \E)$ via $\alpha$.''

As Shimojo and Ichikawa have done, a researcher may assume nothing at
all about a subject's beliefs, except that those beliefs conform to
the model of evidence that is communicated in the experimental
protocol. That situation is represented by $((\Omega, \O, P, \E),
\S)$, where $(\Omega, \O, P, \E)$, is communicated in the protocol and
$\S$ comprises \emph{all\/} of the pairs, $(\Phi, \B, Q, \Omega, \E, \type),
\alpha)$, such that $(\Phi, \B, Q, \Omega, \E, \type)$ conforms to $(\Omega,
\O, P, \E)$ via $\alpha$. Such a situation will be called \emph{full}.

\Section{6}{Characterizing the types of situation}

In terms of the definition of a situation just given, the trichotomy
discussed in the introduction is formalized by
\display{36}{((\Omega, \O, P, \E), \S) \text{\ is of type\ }
\begin{cases}
{1} &\text{if no model of beliefs in $\S$}\\
&\strut\qquad \text{justifies $(\Omega, \O, P, \E)$}\\
{2} &\text{if every model of beliefs in $\S$}\\
&\strut\qquad \text{justifies $(\Omega, \O, P,
  \E)$}\\
{3} &\text{otherwise}
\end{cases}}

If $(\Omega, \O, P, \E)$ is a model of evidence, then define $\E'$ to
be an \emph{almost sure partition} iff, for every pair of distinct 
elements, $C$ and $D$, of $\E'$, $P(C \cap D) = 0$.

\propo{2}{Let $((\Omega, \O, P, \E), \S)$ be a situation. Then
\begin{enum}
\meti{37}
{\strut\qquad If situation $\S$ is not balanced, then it is of type 1.}
\meti{38}
{\strut\qquad If situation $S$ is full and of type 1, then it is not balanced.}
\meti{39}
{\strut\qquad If $\E'$ is an almost sure partition, then situation $\S$ is of type 2.}
\meti{40}
{\strut\qquad If situation $\S$ is full and of type 2, then $\E'$ is an
  almost sure partition.}
\meti{41}
{\strut\qquad There exist situations of each of the three types.}

\end{enum} 
}

\begin{proof}Consider each of the claims.

\noindent[\eqnref{37}]\indent
{This follows from \prp{1}.}

\noindent[\eqnref{38}]\indent
Equivalently, if situation $\S$ is full and balanced, then it is not
of type 1. Assume the antecedent. Because the situation is balanced, it
reflects some model of beliefs. Because the situation is full, that
model of beliefs is in $\S$. That is, the situation is not of type 1.

\noindent[\eqnref{39}]\indent
Suppose that $\E'$ is an almost sure partition and that $((\Phi, \B, Q,
\E, \type), \alpha) \in \S$. Since $(\Phi, \B, Q, \Omega, \E, \type)$ conforms
to $(\Omega, \O, P, \E)$, condition \eqn{18} holds. Together with the
fact that $\E'$ is an almost sure partition, \eqn{18} implies that, for
every $B \in \E'$, $Q(\alpha(B) \triangle \beta(B)) = 0$. It follows
that $(\Omega, \O, P, \E)$ reflects $(\Phi, \B, Q, \Omega, \E,
\type)$. Thefore the situation is of type 2.

\noindent[\eqnref{40}]\indent
Equivalently, if the situation is full and $\E'$ is not an almost sure
partition, then the situation is not of type 2. That is, in that case,
there is some model of beliefs, $(\Phi, \B, Q, \Omega, \E, \type)$, in $\S$
that does not reflect $(\Omega, \O, P, \E)$.  Since $\E'$ is not an
almost sure partition, there are two distinct elements of $\E'$, $C$
and $D$, such that $P(C \cap D) > 0$. 

If the situation is not balanced, then it is of type 1 by
\eqn{37}, so assume that it is balanced. Let $\theta$ be a
balancing function. A model of evidence that conforms to
$(\Omega, \O, P, \E)$, but that does not justify $(\Omega, \O,
P, \E)$, will be constructed.  To do so, let $\Phi$, $\B$,
$Q$, $\type$, and the enumeration of $\E'$, $\langle B_s
\rangle_{s \in S}$, be as in the proof of \lem{3}. (Again,
suppose that $S = \{ 1,\dotsc, n \}$ if $\E'$ has $n$ elements
and that $S = \setbrac{1,2,3\dots}$ if $\E'$ is infinite.)
Without loss of generality, assume that $P(B_1 \cap B_2) > 0$
and that $B_2 \not \subseteq B_1$. By \eqn{7}, $P(B_2
\setminus B_1) > 0$.  Since the situation is full, the
situation to be constructed is in $\S$, so the situation
cannot be of type 2.

To construct the model of beliefs, a type function, $\sigma
\from \Phi \to \E'$, will be constructed. Begin by noting that
$Q(\Omega \times [0, 1/2)) = 1/2$. On $\Omega \times [0,
    1/2)$, $\sigma$ will satisfy $\sigma(\omega, \psi) =
    \tau(\omega, 2\psi)$. This specification ensures that, for
    each $B \in \E'$, $0 < Q(\sigma^{-1}(B)) < 1$ as required
    by condition \eqn{13} of the definition of a model of
    beliefs. On $\Omega \times [1/2,1)$, define $\sigma$ to
      make $\beta(B_1)$ as large as condition \eqn{18} will
      permit

Specifically, define $\mu(\omega) = \min \setbrac{ s \mid
  \omega \in B_s}$, and define $\sigma$ by
\display{42}{\sigma(\omega, \psi) = \begin{cases} \tau(\omega,
    2\psi) & \text{if\ } \psi < 1/2\\
    B_{\mu(\omega)} & \text{otherwise}
    \end{cases}}
With $\beta(B) = \type^{-1}(B)$ and $\gamma(B) =
\sigma^{-1}(B)$, note that
\display{43}{\begin{aligned}
Q(\gamma(B_2)) &= Q((\gamma(B_2) \cap (\Omega \times \left[
      0,1/2 \right))) \cup ((\gamma(B_2) \cap
      (\Omega \times \left[ 1/2,1 \right)))))\\
&= Q((\gamma(B_2) \cap (\Omega \times \left[
      0,1/2 \right)))) + Q(((B_2 \setminus B_1) \times
      (\Omega \times \left[ 1/2,1 \right))))\\
        & > Q((\gamma(B_2) \cap (\Omega \times \left[ 0,1/2
          \right))))
\end{aligned}}
and
\display{44}{\alpha(B_1 \cap B_2) \cap \gamma(B_2) =
  \alpha(B_1 \cap B_2) \cap (\gamma(B_2) \cap (\Omega \times
  \left[ 0,1/2 \right)))}
  Also note that, for all $A \in \O$ and $B \in \E'$,
  \display{45}{Q(\alpha(A) \cap (\gamma(B) \cap (\Omega \times
  \left[ 0,1/2 \right)))) = \frac{Q(\alpha(A) \cap \beta(B))}{2}}
  (and specifically, taking $A = \Omega$, $Q(\gamma(B) \cap (\Omega \times
  \left[ 0,1/2 \right))) = Q(\beta(B))/2$).
Then
\display{46}{\begin{aligned}
    Q[\alpha(B_1 \cap B_2) | \gamma(B_2)] &=
    \frac{Q(\alpha(B_1 \cap B_2) \cap
      \gamma(B_2))}{Q(\gamma(B_2))}\\
    &< \frac{Q(\alpha(B_1 \cap B_2) \cap (\gamma(B_2) \cap
      (\Omega \times \left[ 0,1/2 \right))))} {Q(\gamma(B_2)
        \cap (\Omega \times \left[ 0,1/2 \right)))}\\
  &= \frac{Q(\alpha(B_1 \cap B_2) \cap
          \beta(B_2))}{Q(\beta(B_2))}\\
  &= Q[\alpha(B_1 \cap B_2) | \beta(B_2)]
\end{aligned}}
By the construction of $(\Phi, \B, Q, \Omega, \E, \tau)$ in
\lem{3}, $Q[\alpha(B_1 \cap B_2) | \beta(B_2)] = P[B_1 \cap
  B_2 | B_2]$. Therefore, by \eqn{46}, $Q[\alpha(B_1 \cap B_2)
  | \gamma(B_2)] < P[B_1 \cap B_2 | B_2]$. That is, $(\Phi,
\B, Q, \Omega, \E, \sigma)$ conforms to $(\Omega, \O, P, \E)$
but it does not justify $(\Omega, \O, P, \E)$. Since situation
$\S$ is full, $(\Phi, \B, Q, \Omega, \E, \sigma) \in
\S$. Therefore, situation $\S$ is not of type 2.

\noindent[\eqnref{41}]\indent
By \lem{1}, there is a situation, and hence a full situation,
corresponding to every model of evidence. There are models of evidence
that are not balanced, so, by \eqn{37}, there are full situations of
type 1. There are models of evidence such that $\E'$ is an almost sure
partition, so, by \eqn{39}, there are situations of type
2. \uppercase\xmpl{3} is a balanced model of evidence, for which $\E'$
is not an almost sure partition, so, by \eqn{38} and \eqn{40}, the
full situation corresponding to that model of evidence is of type 3.
\end{proof}

\Section{7}{Conditioning expected utility on evidence and beliefs}

\citet{ShimojoIchikawa-1989} elicited subjects' reports of their
posterior beliefs. They analyzed that data under the assumptions that
(a) their experimental protocol induced specific prior beliefs that
the researchers intended subjects to hold, and (b) subjects were
capable of reporting precise numerical subjective probabilities of
events and were willing to report those probabilities truthfully.

When those assumptions do not hold, another approach must be taken.
One such approach, inspired by the characterization of subjective
utility provided by \citet{Ramsey-1931} and \citet{Savage-1972}, is to
infer subjects' prior and posterior probability measures from data
regarding their choices among alternatives offered in the experiment.

Experimenters studying the Monty Hall problem, such as those
of \citet{GranbergBrown-1995} and \citet{Friedman-1998}, have adopted
a hybrid approach. They have assumed subjects to hold particular prior
probabilities, but have inferred posterior probabilities from observed
choices. In principle, though, some of the reasons to prefer choice-based
imputation of posterior probabilities to subjects should apply to
prior probabilities also. Subjects could be given opportunities to
make choices both before and after having received evidence, with the
former choices revealing information about subjects' prior beliefs and
the latter ones revealing information about posterior beliefs.

Such a thoroughly behavioralistic protocol will be considered now. The
notion of a plan, to be defined momentarily, will play a cognate role
to that of a situation in preceding sections. Essentially, subjects'
choices will be treated as statistics of prior and posterior
beliefs. To observe a statistic of a probability distribution is less
informative than to observe the distribution
directly. Correspondingly, the precise characterization of the various
types of situation in \prp{2} will not have a counterpart
here. Nonetheless, \xmpl{4} will exhibit a plan that can only be
chosen by a subject who reasons according to a model of beliefs,
while \xmpl{6} will exhibit a plan that can only be
chosen by a subject who reasons according to a model of evidence.

\Subsection{7.1}{Plans and conditional expected utility}

Consider an agent who may choose from a set, $\A$, of
alternatives. Suppose that the set of feasible alternatives does not
depend on the state of nature. Let $\E$ be a set of evidential
events. (As specified in \sec{3.1}, $\Omega \in \E \subseteq \O
\setminus \{ \emptyset \} \subseteq 2^\Omega$.)  Intuitively, a plan
is a correspondence that assigns a non-empty set of alternatives to
each evidential event.

A question that it would be typical to pose in decision theory is:
what are the conditions under which a plan, $\plan \from \E \To \A$,
may represent an agent's choices according to maximization of
conditional expected utility? That is, when can $\E$ be associated
with a probability space, and can state-contingent utilities be
imputed to the various alternatives, such that for each $B$,
$\plan(B)$ is the set of alternatives that maximize expected utility
conditional on $B$?

However, regarding the decisions of experimental subjects and of other
agents, there are really two questions. The probability space with
respect to which conditional probabilities are formed might be taken
to be either a model of evidence or else a model of beliefs. If a
model of evidence, then the agent conditions on the event itself. If a
model of beliefs, then the agent conditions on the distinct event,
$\beta(B)$, that the evidential event, $B$, is the agent's type.

The two questions are formulated explicitly as follows.
\begin{enum}
\renewcommand{\theequation}{\arabic{enumi}.\arabic{equation}} \setcounter{equation}{0}
\meti{47}
{Under what conditions on $(\Omega, \O, \E)$ and $\plan$ do there exist a
  probability space, $(\Psi, \C, P)$, an isomorphism $\alpha \from \O
  \to \C$, and a set of (bounded, $\O\/$-measurable,
  state-contingent) utility functions, $\langle u_a \from \Psi \to
  \Re \rangle_{a \in \A}$, such that, for all $a\in A$ and $B \in \E$,

{ \setlength{\parindent}{2pt} \indent\parbox{10cm}{
\display{47.1}{\int_{\alpha(B)} u_a \, dP = \max_{b \in \A}
  \int_{\alpha(B)} u_b \, dP \iff a \in \plan(B)}}
}\\
That is, under what conditions do there exist a model of evidence,\\
$(\Psi, \C, P, \alpha(\E))$, and a set of utility functions that
rationalize $\plan$?}
\setcounter{equation}{0}
\meti{48}
{Under what conditions on $\E$ and
  $\plan$ do there exist a model of beliefs,\\ $(\Phi, \B, Q, \Omega, \E,
  \type)$ and a set of bounded utility functions,
  $\langle v_a \from \Phi \to \Re \rangle_{a \in \A}$, such that, for
  all $a \in \A$ and $B \in \E$, 

{ \setlength{\parindent}{2pt} \indent\parbox{10cm}{
\display{48.1}{\int_{\beta(B)} v_a \, dQ = \max_{b \in \A}
    \int_{\beta(B)} v_b \, dQ \iff a \in \plan(B)}
}}
}
\end{enum}
\setcounter{equation}{\value{enumi}}
\renewcommand{\theequation}{\arabic{equation}}

A model of evidence that satisfies the condition stated in \eqn{47}
\emph{rationalizes $\plan$ by evidence}. A model of beliefs that
satisfies the condition stated in \eqn{48} \emph{rationalizes $\plan$
  by beliefs}. A plan that is rationalized by some model is called
\emph{rational} with respect to that type of model.

The definition of rationalization by a model of beliefs shows why the
prior-beliefs type, $\Omega$, is needed although it is not in the
range of the type function. If the definition of $\E$ were
amended so that $\Omega \notin \E$, then any plan could be
rationalized by beliefs.
The reason is that, since $\setbrac{\beta(B) \mid B \in \E'}$
is a partition of $\Phi$, functions $v_a$ can be defined by
\display{49}{v_a(\phi) = \begin{cases}
1 & \text{if\ } a \in \plan(\type(\phi))\\
0 & \text{otherwise}
\end{cases}}
However, because the definition of rationality with respect to beliefs
requires that \eqn{48} must be satisfied also by $B=\Omega$, \eqn{49}
does not automatically define utility functions that rationalize
$\plan$. This observation reflects the basic principle that the force of
Bayesian decision theory comes from the relationship between choices
based on prior versus posterior beliefs, not solely on relationships
among choices conditioned on alternate posterior beliefs.

\uppercase\sec{7.2} concerns an example of a plan that is
rational with respect to beliefs, but not with respect to
evidence.  \uppercase\sec{7.3} concerns an example of a plan
that is rational with respect to evidence, but not with
respect to beliefs. In fact, this example formalizes the Monty
Hall problem that has been studied in various experiments
cited earlier.

\Subsection{7.2}{Rationality with respect to beliefs does not imply
  rationality with respect to evidence}

In this section, first a model of beliefs will be constructed and will
be shown to rationalize a plan. Then, it will be shown that no model
of evidence can rationalize that plan.

\exmpl{4}{ The model of beliefs closely resembles \xmpl{2}. It is based
  on a model of evidence that differs from \xmpl{1} by the addition of
  a new evidential event, $\half$. Thus, let $\Omega = \{e,h,f \}$ and
  $\E = \{ \Omega, \mpty, \half, \full \}$, where $\mpty = \{ e,h \}$,
  $\full = \{ h,f \}$, and $\half = \setbrac{h}$.  Besides the addition of
  $\half$ to $\E$, the other change of the current example from from
  \xmpl{2} is to put greater weight on $h$ than is specified in that
  earlier example. The set of states of the world, $\Phi$, of the model of beliefs
  will be $\Omega \times \E'$, and prior beliefs will be specified so
  that $Q( \setbrac{h} \times \E') = 3/5$ and $Q( \setbrac{f} \times \E' )= Q(
  \setbrac{e} \times \E' ) = 1/5$.

The specification of the model of beliefs is completed by taking $\B =
2^\Phi$, and $\tau(\omega, B) = B$, and by fully specifying $Q$
according to the following table. The cells are interpreted as in
table \eqn{15}.
\display{50}{ \begin{tabular}[c]{|c||c|c|c|c|}
\hline
$Q(\omega,B)$ & $\Omega$ & $\mpty$ & $\half$ & $\full$\\

\hline \hline

$e$ & .2 & .2 & 0 & 0\\

\hline

$h$ & .6 & .1 & .4 & .1\\

\hline

$f$ & .2 & 0 & 0 & .2\\

\hline

\end{tabular} }

Since $\type(\omega, B) = B$, $\beta(B) = B \times \E'$. Specify $\A$
by $\A = \{ \w, \d \}$.  Specify that $\plan(\Omega) = \plan(\half) =
\{ \w \}$ and $\plan(\mpty) = \plan(\full) = \{ \d \}$.}

\claim{11}{The plan specified in \xmpl{4} is rational with respect to beliefs, but
  not with respect to evidence.}

\begin{proof}
It will be shown that there are utility functions that, together with
the model of beliefs constructed in \prp{1}, rationalize $\plan$ in
\xmpl{4}. However, $\plan$ is not rational with respect to evidence.

Intuitively,  $\w$ is supposed to be
wagering that the state of nature is $h$ and $\d$ is supposed to be declining to
wager. Formally suppose that, for all $B \in \E'$, $v_\w(h,B) = 10$ and
$v_\w(e,B) = v_\w(f,B) = -10$ and that, for all $\omega \in \Omega$,
$v_\d(\omega) = 0$. Then
\display{51}{\begin{gathered}
\int_{\beta(\Omega)} v_\w - v_\d \; dQ = 2 \quad \text{and} \quad
  \int_{\beta(\half)} v_\w - v_\d \; dQ = 4  \quad \text{and}\\
\int_{\beta(\mpty)} v_\d - v_\w \; dQ = \int_{\beta(\full)} v_\d - v_\w
\; dQ = 1
\end{gathered}}
so $\plan$ is rational with respect to beliefs.

A contradiction will be obtained from supposing that some isomorphism,
$\alpha \from \O \to \C$, model of evidence, $(\Psi, \C, R,
\alpha(\E))$, and set of utility functions, $\langle u_a \from \Psi
\to \Re \rangle_{a \in \A}$, rationalize $\plan$

\display{52}{\begin{aligned}
\plan(\mpty) = \{ d \},\text{\ so\ } &\int_{\alpha(\mpty)} u_\d - u_\w \, dR > 0\\
\plan(\half) = \{ w \},\text{\ so\ } &\int_{\alpha(\half)} u_\d - u_\w \, dR < 0\\
\text{thus\ } &\int_{\alpha(\mpty) \setminus \alpha(\half)} u_\d - u_\w \, dR > 0\\
\plan(\full) = \{ w \},\text{\ so\ } &\int_{\alpha(\full)} u_\d - u_\w \, dR > 0\\
\int_{\alpha(\Omega)} u_\d - u_\w \, dR = &\int_{\alpha(\full)} u_\d - u_\w \, dR +
\int_{\alpha(\mpty) \setminus \alpha(\half)} u_\d - u_\w \, dR   > 0
\end{aligned}}

But, that $\int_{\alpha(\Omega)} u_\d - u_\w \, dR > 0$ and $\plan(\Omega) = \{
\w \}$ contradicts condition \eqn{48.1} for $\alpha$, $(\Psi, \C, R,
\alpha(\E))$, and $\langle u_a \rangle_{a \in \A}$ to rationalize $\plan$.
\end{proof}

\Subsection{7.3}{Rationality with respect to evidence does not
  imply rationality with respect to beliefs}

The BFG problem is in what \citet{Savage-1972} called the
``verbalistic'' tradition, while the Monty Hall (MH) problem is in the
``behavioralistic'' tradition. That is, the BFG problem is formulated
in terms of eliciting first-person reports of an agent's probability
assessments, while the MH problem is formulated in terms of acquiring
evidence about the pattern of the agent's practically significant
decisions. In a situation where it is possible to observe an agent's
choices but not to query the agent about probability assessments, or
where it is thought that an agent's responses to such queries will
either over- or under-state the agent's capacity to act in conformity
to expected-utility maximization, the MH problem could be the more
advantageous one to consider.

Of course, the BFG problem can be reformulated in a behavioralistic
framework. This will be done now. It will be shown that the plan that
corresponds naturally to heuristic reasoning is rational with respect
to beliefs, as well as with respect to evidence. Thus, in the
situations just envisioned, the BFG problem cannot be used to design
an experiment, the outcome of which could rule out the possibility
that an agent reasons soundly according to a model of beliefs. The
plan that corresponds naturally to heuristic reasoning in the MH
problem is defined from the same evidential events and alternatives as
is the previous plan. That plan is rationalized by the model of
evidence presented in \xmpl{1}, amended so that the prior probabilities
are modified so that each is $1/3$, along with the same utility
functions by which the heuristic plan for the BFG problem is
rationalized. However, it will be shown that the heuristic plan for
the MH problem is not rational with respect to evidence.

Consider the behavioralistic formulation of the BFG problem.

\exmpl{5}{Specify $\Omega$, $\O$, and $\E$ as in \xmpl{1}, and let
$\A = \{ \e, \h, \f \}$. Specify that $\plan(\Omega) = \plan(\mpty) =
  \plan(\full) = \{ \h \}$.}

Define $\a \from \Omega \to \A$ by $\a(e)
= \e$, $\a(h) = \h$, and $\a(f) = \f$. For $\omega \in \Omega$ and $b
\in \A$, specify that
\display{53}{u_b(\omega) = \begin{cases}
1 &\text{if\ } b = \a(\omega)\\
0 &\text{if\ } b \neq \a(\omega)
\end{cases}}
In \xmpl{1}, $P( \setbrac{h} ) = 2/5$ and $P( \setbrac{e} ) = P( \setbrac{f} ) =
3/10$. Consequently\\ $P[ \setbrac{h} | \mpty ] = P[ \setbrac{h} | \full ] =
4/7$. Thus the model of evidence in \xmpl{1}, together with the
utility functions defined in \eqn{53}, rationalize $\zeta$ with
respect to evidence. That is, on the intuitive understanding of the
alternatives that was suggested above, $\zeta$ is the plan that
corresponds naturally to heuristic reasoning in the BFG problem.

Plan $\plan$ is also rational with respect to beliefs. One way of
showing that is to appeal to the model of beliefs constructed in \xmpl{2},
and to specify that, for all $\phi \in \Phi$, $v_\e(\phi) =
v_\f(\phi) = 0$ and $v_\h(\phi) = 1$. Rationality with respect to
beliefs can also be shown by defining $v_b(\omega, B) = u_b(\omega)$
and by modifying $Q$ from \eqn{15} to the following specification, which
assigns very high probability to the event that $h$ is the state of nature.
\display{54}{ \begin{tabular}[c]{|c||c|c|c|}
\hline
$Q(\omega,B)$ & $\Omega$ & $\mpty$ & $\full$\\

\hline \hline

$e$ & .1 & .1 & 0\\

\hline

$h$ & .8 & .4 & .4\\

\hline

$f$ & .1 & 0 & .1\\

\hline

\end{tabular} }
This second way has the feature that the alternatives continue to be
given their intuitive interpretations according to the assignment of
utilities, rather than $\h$ being treated as a dominant alternative.

In contrast, the way that subjects are understood to reason in Monte
Hall experiments suggests a plan that is rational with respect to
evidence, but is not rational with respect to
beliefs.\footnote{\citet[pp.\ 711, 712]{GranbergBrown-1995}
  hypothesize that their subjects reason heuristically as in \xmpl{1},
  in a setting tantamount to that example except
  that the prior probability of each state of nature is $1/3$. This prior
  makes the two alternatives consistent with the subject's type to be
  equal to one another in expected utility. Subjects cannot express
  indifference, given the forced-choice protocol of the
  experiment. Granberg and Brown suggest that ``intertia'' or some
  other tie-breaking consideration accounts for subjects' expressed
  choices, implying that those choices represent a single-valued
  selection from an underlying plan that is a multi-valued
  correspondence. They hypothesize that some subjects may, in fact, be
  randomizing between the two alternatives.} In its original form, the
MH problem involves an agent making a provisional choice, and
subsequently having an opportunity to revise that choice. The
specification of of $\E'$ in \xmpl{1}, which is incorporated in 
following example, corresponds to the revised-choice stage of the MH
problem that would follow the agent having made $\h$ as the
provisional choice.

\exmpl{6}{The example is identical to \xmpl{5}, except that
\display{55}{\plan(\Omega) = \A \qquad \plan(\mpty) = \{ \e, \h \}
\qquad \plan(\full) = \{ \h, \f \} }
}

\claim{14}{The plan specified by \eqn{55} in \xmpl{6} is rational
  with respect to evidence, but not with respect to beliefs.}

\begin{proof}
Specify a model of evidence according to \xmpl{6}, along with the
specification that $P(\e) = P(\h) = P(\f) = 1/3$. This model and the
utility functions defined by \eqn{53} rationlize $\zeta$.

Now it will be shown by contradiction that there do not exist a model
of beliefs, $(\Phi, \B, Q, \Omega, \E, \type)$ and utility functions $\langle
v_b \rangle_{b \in \A}$ that rationalize $\zeta$. 
By \eqn{48.1}, since $\h \in \plan(\Omega) \cap \plan(\mpty)$
and $\f \in \plan(\Omega) \setminus \plan(\mpty)$,
\display{56}{\int_{\beta(\Omega)} v_\h \, dQ = \int_{\beta(\Omega)} v_\f
  \, dQ}
and
\display{57}{\int_{\beta(\mpty)} v_\h \, dQ > \int_{\beta(\mpty)} v_\f
  \, dQ}
Since $\beta(\Omega) = \Omega$, \eqn{56} implies that
\display{58}{\int_\Omega v_\h \, dQ = \int_\Omega v_\f \, dQ}
Because $\E' = \{ \mpty,\full \}$, $\{ \beta(\mpty), \beta(\full) \}$
is a partition of $\Omega$. Therefore, \eqn{57} and \eqn{58} imply that
\display{59}{\int_{\beta(\full)} v_\f \, dQ > \int_{\beta(\full)} v_\h \, dQ}
But, given condition \eqn{48.1}, inequality \eqn{59} contradicts a clause of
assumption \eqn{55}, that $\h \in \plan(\full)$.
\end{proof}

%\bibliographystyle{plainnat}
%\bibliography{heuristic}

\begin{thebibliography}{15}
\providecommand{\natexlab}[1]{#1}
\providecommand{\url}[1]{\texttt{#1}}
\expandafter\ifx\csname urlstyle\endcsname\relax
  \providecommand{\doi}[1]{doi: #1}\else
  \providecommand{\doi}{doi: \begingroup \urlstyle{rm}\Url}\fi

\bibitem[Aliprantis and Border(2006)]{AliprantisBorder-2006}
Charalambos~D. Aliprantis and Kim~C. Border.
\newblock \emph{Infinite Dimensional Analysis: A Hitchhiker's Guide}.
\newblock Springer, third edition, 2006.

\bibitem[Bar-Hillel and Falk(1982)]{BarhillelFalk-1982}
Maya Bar-Hillel and Ruma Falk.
\newblock Some teasers concerning conditional probabilities.
\newblock \emph{Cognition}, 11:\penalty0 109--122, 1982.

\bibitem[Bertrand(1889)]{Bertrand-1889}
J.~Bertrand.
\newblock \emph{Calcul des Probabilit{\'e}s}.
\newblock Gauthier-Villars, 1889.
\newblock URL \url{https://books.google.com/books?id=IDo7AQAAIAAJ}.

\bibitem[Friedman(1998)]{Friedman-1998}
Daniel Friedman.
\newblock Monty \uppercase{H}all's three doors: construction and deconstruction
  of a choice anomaly.
\newblock \emph{American Economic Review}, 88:\penalty0 933--946, 1998.

\bibitem[Friedman(1953)]{Friedman-1953}
Milton Friedman.
\newblock The methodology of positive economics.
\newblock In Milton Friedman, editor, \emph{Essays in Positive Economics}.
  University of Chicago Press, 1953.

\bibitem[Gardner(1959)]{Gardner-1959}
Martin Gardner.
\newblock Mathematical games.
\newblock \emph{Scientific American}, 201:\penalty0 (4):174--182, (5):181--188,
  1959.

\bibitem[Granberg and Brown(1995)]{GranbergBrown-1995}
Donald Granberg and Thad~A. Brown.
\newblock The \uppercase{M}onty \uppercase{H}all dilemma.
\newblock \emph{Personality and Social Psychology Bulletin}, 21:\penalty0
  711--723, 1995.

\bibitem[Harsanyi(1967)]{Harsanyi-1967}
John~C. Harsanyi.
\newblock Games with incomplete information played by ``\uppercase{B}ayesian''
  players, \uppercase{I}-\uppercase{III} part \uppercase{I}. the basic model.
\newblock \emph{Management Science}, 14\penalty0 (3):\penalty0 159--182, 1967.

\bibitem[Kluger and Friedman(2010)]{KlugerFriedman-2010}
Brian Kluger and Daniel Friedman.
\newblock Financial engineering and rationality: experimental evidence based on
  the \uppercase{M}onty \uppercase{H}all problem.
\newblock \emph{The Journal of Behavioral Finance}, 11:\penalty0 31--49, 2010.

\bibitem[Ramsey(1931)]{Ramsey-1931}
Frank~Plumpton Ramsey.
\newblock Truth and probability.
\newblock In R.~B. Braithwaite, editor, \emph{Foundations of mathematics and
  Other Logical Essays}. Routledge, 1931.

\bibitem[Savage(1972)]{Savage-1972}
Leonard~Jimmie Savage.
\newblock \emph{The Foundations of Statistics}.
\newblock Dover, second revised edition, 1972.

\bibitem[Selvin(1975)]{Selvin-1972a}
Steve Selvin.
\newblock A problem in probability.
\newblock \emph{The American Statistician}, 29:\penalty0 67, 1975.

\bibitem[Shimojo and Ichikawa(1989)]{ShimojoIchikawa-1989}
Shinsuko Shimojo and Shin'ichi Ichikawa.
\newblock Intuitive reasoning about probability: theoretical and experimental
  analysis of the `problem of three prisoners'.
\newblock \emph{Cognition}, 32:\penalty0 1--24, 1989.

\bibitem[Sikorski(1969)]{Sikorski-1969}
Roman Sikorski.
\newblock \emph{Boolean Algebras}.
\newblock Springer, third edition, 1969.

\bibitem[Simon(1955)]{Simon-1955}
Herbert~A. Simon.
\newblock A behavioral model of rational choice.
\newblock \emph{Quarterly Journal of Economics}, 69:\penalty0 99--118, 1955.

\end{thebibliography}

\strut \end{document}